# Study of impurity distribution in mechanically polished, chemically treated and high vacuum degassed pure Niobium samples using TOFSIMS technique

A. Bose [a], S. C. Joshi [a]

[a] Proton Linac and Superconducting Cavities Division

 *Raja Ramanna Centre for Advanced Technology, Indore, M. P. 452013, India*

## Abstract

The performance of Superconducting radio frequency cavities (SRF) is strongly influenced by various impurities within the penetration depth (~50nm) of Nb, which in turn depends on the applied surface treatments. The effect of these surface treatments on the impurities of Nb has been explored using various surface analytical treatments. But, the results are still inadequate in many aspects and the effect of sequential SRF treatments on the impurity distribution has not been explored. The present study analyses various impurities within the penetration depth of Nb samples, treated by SRF cavity processing techniques like colloidal silica polishing (simulating centrifugal barrel polishing), buffer chemical polishing (BCP), high pressure rinsing (HPR) and degassing under high vacuum (HV) condition at 600°C for 10hrs. Static, dynamic and slow sputtering modes of Time of flight secondary ion mass spectrometry (TOFSIMS) technique was employed to study the effect of the above treatments on interstitial impurities, hydrocarbons, oxides, acidic residuals, reaction products and metallic contaminations. The study confirms that the impurity distribution in Nb is not only sensitive to the surface treatment, but also to their sequence. Varying the treatment sequence prior to HV degassing treatments affected the final impurity levels in HV degassed bulk Nb samples. BCP treated samples, exhibited minimum hydrocarbon and metallic contamination but, led to extensive contamination of the oxide layer with residuals and reaction products of acids used in BCP solution. HPR treatment, on the other hand was effective in reducing the acidic impurities on the top surface. The study also establishes the application of TOFSIMS technique to analyze and evolve SRF treatments.

# 1. Introduction

Superconducting radio frequency (SRF) cavities made out of high purity niobium (Residual resistivity ratio, RRR ~ 300) is the primary choice for the next generation linear accelerators (Linac). Research and development activities for building a 1 GeV superconducting linac (Indian spallation neutron source) has been initiated at RRCAT, Indore [1]. The medium and high energy section of this linac would consist of 650 MHz SCRF cavities made of Nb. The optimized target gradient of these cavities would be ~19 MV/m at proposed quality factor of ~$10^{10}$ [2] and should have minimum field emission (FE) and multipacting. The optimize performance of the cavities depends on the established sequence of standard chemical and thermal treatments. For example, the average accelerating gradient of 1.3GHz electropolished (EP) cavities were ~15% higher than buffer chemical polished (BCP) cavities [3]. Moreover, low temperature baking (LTB), improves the gradient and $Q_O$ in EP as well as BCP treated cavities [4]. But, in some exceptional cases, stand-alone BCP treated 1.3GHz cavities have also reached 40MV/m, which adds complexity to the above subject [5]. Since large numbers of cavities are required to build a superconducting accelerator, so these treatments should ensure repeatability in their performance. The primary effect of these treatments is to alter the impurities within the RF penetration depth of Nb, which is close to ~50 nm [6]. It was found that, if the phosphoric acid in BCP solution is replaced with sulphuric acid, the Q-slope improves in comparison to the standard BCP treated cavity [5], which further highlights the role of impurities. Various forms of impurities that include interstitials [7-8], metallic, non-metallic [9-10], hydrocarbons [11-12] and oxides [10,13-18]; combined with their locations that includes, surface [9-11,19-21], subsurface [7-8,14,16-17,22], inside penetration depth [7,13], bulk [4] and grain boundaries [15,23]; were directly or indirectly linked to the cavity performance [7-18]. These impurities indirectly affect superconducting parameters like critical temperature ($T_c$) [24-25], the upper critical field ($H_{c2}$) [25], surface superconductivity [11] and residual resistance [19] of Nb, which in turn might affect the maximum gradient and the Q-slope. Hence, the role of impurities in inducing variation in cavity performance assumes importance; which forms the basis of present research. The study critically analyses the distribution of various impurities in Nb after standard SRF treatments using Time of flight secondary ion mass spectrometry (TOF-SIMS 5 from M/s IONTOF, GmbH) technique. The effect of SRF treatments on few impurities have been analyszed using techniques like XPS [14,21,22,26], AES [27-28], ERDA [8], TEM [17,29-31], Raman [12], SEM-EDX [32] and SIMS [13,21,33-34]. But, detailed study of the total spectrum of impurities is still lacking.

TOFSIMS technique was selected owing to its quasi parallel detection capability, surface sensitivity, low detection limit combined with large coverage of elements. Moreover, high lateral and depth resolution ensures analysis with greater details, which can be complemented with the results from other techniques. Earlier investigations using SIMS have focused on the oxide layer development against oxygen exposure [35-36], while others have proposed mechanism to investigate the oxide composition [33,37]. Studies have also concentrated on the effect of high vacuum thermal degassing (HV degassing) and LTB on interstitial impurities [7,34]. The study demonstrates the role of $H_2$ partial pressure on the level of H reduction [7]. But, the role of SRF treatments prior to HV degassing was not investigated. Therefore, limited emphasis has been placed to understand the step-wise effect of SRF treatments on the vast spectrum of impurities that include interstitials, hydrocarbons, oxides, acidic residual, reaction products and metallic. The present investigation uses different modes of TOFSIMS to analyze the changes in impurities present on surface, subsurface as well as penetration depth (~50nm), due to sequential treatments of colloidal silica polishing (CSP), BCP, high pressure rinsing (HPR) and thermal degassing treatment. Previous studies have also reported sputtering of intense $NbH_x$ (x=1-5) fragments during H depth profiling [7, 38]. But, attempt to explain the origin was absent. The present study not only develops the methodology to analyze the effect of SRF treatments using TOFSIMS, but, also tries to explain the origin of most sputtered ionic species from impurities present inside Nb. The purpose of choosing CSP as baseline treatment stems from two factors. First, the surface condition of the samples, cut from the raw Nb sheets were unknown. Second, CSP is the proposed final step in centrifugal barrel polishing treatment, prior to chemical treatments to cavities being processed at RRCAT. Finally selection of BCP treatment was based on two prime reason. First, BCP treated 650MHz cavities have shown encouraging results at Jefferson Lab [39] and second is the rise of large grain cavities. The study would be useful in analyzing the complex EP process in future which has more variables like voltage, temperature, anode cathode ratio, anode cathode distance etc.

## 2. Experimental

### 2.1. Materials

In this work four high purity Niobium samples P78, P79, P115 and P117 of 13mm x 8mm (approx.) sizes, were cut from polycrystalline niobium sheet (2.8mm thick) procured from M/s Plansee GmbH. The physical properties and chemical composition of the niobium samples are mentioned in table 1.

### 2.2. Preparation

The raw samples were sequentially cleaned by ultrasonic rinsing in Micro-90 (M90) detergent solution and ultra-pure water (UPW). The step-wise treatments on two sets of raw and cleaned samples (P78 and P79) are listed in table 2, which focused on following the SCRF cavity processing sequence. The raw and cleaned samples were initially rough polished on silicon carbide papers and fine polished using diamond paste, followed by colloidal silica polishing (CSP) using colloidal solutions dispersed with 40 nm silica particles. It was again cleaned by sequential ultrasonic rinsing in M90 solution and UPW for 30 minutes each and then analyzed using TOFSIMS technique. These samples were identified as P78C and P79C. These samples (P78C and P79C) were subsequently BCP treated for at least 60 minutes in a fresh acid mixture of HF: $HNO_3$: $H_3PO_4$ (1:1:2), stirred at a constant speed using teflon coated magnetic stirrer. The temperature of the BCP solution was monitored using a Teflon coated thermocouple and maintained within 11°C to minimize hydrogen uptake [40]. The average etching rate was ~0.65 μm/min. These samples were ultrasonically rinsed in UPW for 60 minutes and identified as P78B and P79B. Finally, they were high pressure rinsed (HPR) in HPR-set up inside class 100 clean room. The samples were HP rinsed for 10 minutes at a distance of ~10 cm from the HPR nozzle using a sample holder arrangement. The water pressure was maintained at 85 kg/cm². These samples were identified as P78H and P79H.

Another two sets of sample, P115 and P117, were BCP treated and CS polished respectively. These samples were analyzed and subsequently annealed at 600°C for 10hrs in high vacuum annealing furnace (< 2 x 10⁻⁶ mbar) and identified as P115U and P117U. The twin purpose of these two samples was to investigate effect of HV degassing treatment and the role of treatment sequence on the impurity distribution.

Processing and handling of all samples were done with extreme precaution to avoid any external contamination apart from the treatment step itself. The sample surface was kept untouched and loaded inside SIMS loadlock chamber (< 3x 10⁻⁷ mbar) within 30 minutes of the concerned preparation step except P115 and P117.

The above procedure of treatments and analysis were repeated again on other Nb samples including samples from a different supplier to ensure repeatability and the results exhibited similar trend. Any variations that were observed have also been incorporated in the present article.

## 2.3. TOF-SIMS

The measurements using TOFSIMS technique were divided into three parts. Part 1 carried out impurity analysis of the top surface using Static SIMS technique; Part 2 was devoted to oxide layer analysis using slow sputtering technique and Part 3 performed impurity analysis throughout the penetration depth using dynamic SIMS technique.

The static SIMS spectrum of the two sets of samples (P78 and P79), after each treatment, were obtained in positive and negative secondary ion mode using a pulsed 30keV $Bi_1^+$ primary ion beam at ~2pA current, rastered over an area of 350µm x 350µm. To ensure analysis within the static limit, the primary ion dose was kept within 7 x $10^{11}$ions/cm$^2$. The spectra were recorded at a minimum of 3 fresh spots on each sample to measure repeatability. The counts of each impurity from 3 spots were averaged and subsequently compared after each surface treatments. The detailed procedure of normalization repeatability measurements shall be explained in section 3.1.

The second stage of analysis involved slow sputtering of the oxide layer using pulsed $Bi_1^+$ primary ion beam (30keV, ~4pA) alone for an extended period (>4000sec). The measurements were done after each treatment to compare the oxide layer. The area of analysis was kept at 80µm x 80µm. To remove the edge effect after extended sputtering, the depth profile was reconstructed from the 50µm central area using the TOFSIMS Surface lab software. Single spot was analyzed for each sample. It was ensured that the areas to be analyzed are free from localized impurities. This was confirmed by observing the ion images prior to the oxide layer analysis. These ion images were acquired for 10 seconds in both positive and negative ion mode.

Finally the depth profile analysis up to the penetration depth of Nb was carried out in interlaced mode with pulsed $Bi_1^+$ at 30 keV as the analysis gun and 1keV $Cs^+$ and 1keV $O_2^+$ as the sputter gun operating at a constant current of ~74nA and ~100nA respectively. The area of analysis across all measurements was 100µm x 100µm inside the sputter area of 300µm x 300µm as shown in figure 1. The sputter gun parameters were selected to ensure that erosion is slow enough to reveal the oxide layer and fast enough to cover 50nm depth within a short time to avoid drift in analysis conditions. $Cs^+$ ions improves detection of non-metallic impurities and oxides while $O_2^+$ ion enhances detection of metallic impurities. The depth profiles are plotted in figure 7 and 11, where, X-axis represents time and Y-axis the normalized intensity. The normalization of the counts from negative ion fragments was done with $Nb^-$ counts [7]. Normalization process was performed only after the removal of the top oxide layer, to eliminate the matrix effect. $Nb^-$ intensity was also found to stabilize, after the removal of oxide layer. On the other hand positive ion fragments were compared on the basis of absolute counts, since, proper normalization factor was not available. The depth profiles compare the impurities between samples treated by CS polishing, BCP treatment and subsequent HPR treatment along with BCP treated+HV degassed and CS polished+HV degassed samples respectively. Time on the X-axis was converted to depth by estimating the sputter rate of the sputter gun. The sputter rate estimation requires the measurement of crater depth of Nb created by the sputter guns. Three separate samples were prepared for the above measurement to overcome the crater depth measurement inaccuracies creeping from initial roughness, final crater roughness, sample slope combined with shallow depth created by low energy sputter gun. The samples were i) colloidal silica polished sample (shown in figure 1), ii) ~470 nm Nb thin film grown on Si wafer and iii) set of anodized niobium samples with varying oxide thicknesses. The oxide thickness were calculated on the basis of formula available in literature [16]. The craters of first 2 samples were measured using stylus instrument and 3-D confocal microscope. In case of third sample SIMS sputtering was stopped as we reached bulk Nb. The sputter rate was estimated between 0.175 - 0.24nm/s for 1kV $Cs^+$ and $0.22 - 0.255$nm/s for 1kV $O_2^+$ sputter gun. Average values were used in the depth profiles.

The typical mass resolution was > 8000 at mass $^{28}Si$ for all measurements and vacuum level was less than 5e-10 torr prior to all analysis. To the best of our knowledge we could not find such a detailed analysis which is critical to the understanding of the variability from various treatments applied on the SRF cavities.

## 3. Results and discussions

The results and discussion part was divided in three sections. Section 3.1 analyzes the changes in impurities on the top monolayer using static SIMS approach, while section 3.2 extends the study to the top oxide layer using slow depth profiling with $Bi_1^+$ beam alone and finally section 3.3 reports the impurity analysis up to the penetration depth (~50nm) using dual beam dynamic TOF-SIMS technique.

### 3.1 Top layer impurity analysis using Static SIMS

Analysis of surface specific impurities on Nb is considered essential due to its ability to affect the performance of SRF cavities through FE and multipacting. Many impurities like sulfur [9, 21], aluminium [9], fluorine [21], chlorine [20] and carbon [15,21,32,41] found on EP and/or BCP treated surfaces are considered to be potential sources of field emission (FE) [9,15,20,21,32]. Investigation of FE created craters has also revealed the presence of many non-metallic and metallic impurities [20]. Additionally, multipacting was found to be enhanced due to adsorbates or condensed gas [19]. Assessment of these impurities require extreme surface sensitivity which can be analyzed using Static SIMS approach alone. Moreover, the secondary ion clusters emanated during static SIMS sputtering occurs primarily by atoms that occupy adjacent sites with bonding [42]. Using this approach, the chemical information of BCP treated niobium (P78B) surface was extracted from a typical positive and negative secondary ion spectra shown in Figure 2. The ion fragments from top surface of niobium sputtered $C^{+/-}$, $CH^{+/-}$, $CH_2^{+/-}$ and host of other $C_xH_y^{+/-}$ and $C_xH_yO_z^{+/-}$ ion fragments indicating the presence of hydrocarbons. The hydrocarbons present over the top niobium oxide surface also emitted fragments like $H^{+/-}$, $H_2^{+/-}$, $NbH_x^{+/-}$, $NbC^{+/-}$, $NbCH_x^{+/-}$, $NbC_xH_y^{+/-}$, and $NbO_xC_yH_z^{+/-}$. The fragments sputtered from the top niobium oxide layer includes $NbO^{+/-}$, $NbO_2^{+/-}$, $NbO_3^-$, $NbO_4^-$, $Nb_2O_2^{+/-}$, $Nb_2O_3^{+/-}$, $Nb_2O_4^{+/-}$ and $Nb_2O_5^-$ along with $OH^-$, $NbOH_x^{+/-}$ from hydroxides or HOH terminating on niobium oxides. The distinguishing feature of BCP treated sample was the presence ionic fragments of acidic residues indicated by $F^-$, $F_2^-$, $H_2F^-$, $P^-$, $PO^-$, $PO_2^-$, $PO_3^-$, $N^+$, $NO^{+/-}$, $NOH^-$, $NO_2H^-$, $Cl^-$ as well as reaction products like $NbO_xF_y^-$, $C_xF_y^+$ and $CH_xF^{+/-}$. Metallic impurities indicated by $Na^+$, $Si^{+/-}$, $K^+$, $Ca^+$, $Mg^+$, $Cr^+$, $Ni^+$, $Fe^+$, $Cu^{+/-}$ and fragments indicative of atmospheric contamination like $Cl^-$, $NO_2^-$ and $NO_3^-$ were also observed. This analysis further suggests that smaller fragments like $H^{+/-}$, $C^{+/-}$, $O^{+/-}$ sputtered from various sources that includes, hydrocarbons, amines, oxides, hydroxides, adsorbed water, acidic fragments, reaction products and atmospheric impurities. But, larger fragments were indicators of specific species. Therefore, comparison based on this approach would enable the linkage between contaminations present due to a particular surface treatment with the FE and multipacting observed in SRF cavities.

Prior to comparison of impurities between various treatments, it was necessary to normalize various ion fragments to eliminate errors creeping due to instrumental factors. Nb$^+$ being an abundant species was used for normalization of all the positive ion fragments. But, the process of normalization applies equal weightage to all the ion fragments, which may not be true. As a result the ratios of each ion fragments with Nb$^+$ within a treatment was cross checked. Ion fragments of metallic impurities were found to vary independent of Nb$^+$ counts. So the results of metallic impurities were compared using absolute counts. Similar exercise with negative ion fragments suggested that Nb$^-$ can be chosen as the normalization factor. As we proceed the above choices would obviate into the most probable assumption. After each treatment, the normalized data from three spots of a sample were averaged. Finally, the percentage change of individual species (impurities) of say A$^-$, between CS polishing and BCP treatment was calculated as

$$\% \text{ Change of A}^- \text{ after BCP treatment on CS polished samples} = \left[ \frac{\left( \frac{A^-}{Nb^-} \right)_{BCP} - \left( \frac{A^-}{Nb^-} \right)_{CS}}{\left( \frac{A^-}{Nb^-} \right)_{CS}} \right] \times 100$$

Same procedure was applied for comparing the changes in positive ion fragments. The percentage change of a few negative and positive ion fragments after BCP and subsequent HPR treatment on colloidal silica polished samples (P78C and P79C) are presented in table 3. The ionic fragments in table 3 were divided into four categories that includes hydrocarbons, niobium oxide fragments, acidic residuals/ reaction products and metallic contaminations. The results of reproducibility of each treatment are also presented in section 3.1.5.

3.1.1. Comparison of Hydrocarbons related contamination:

It is clear from table 3, that the hydrocarbons represented by various $C_xH_y^{+/-}$ fragments including C$^-$, H$^-$ reduced by more than 50% after BCP treatment in both set of samples (P78 and P79). Moreover, NbH$_x^{+/-}$, NbC$_x$H$_y^{+/-}$ were also found to decrease after BCP treatment which are also indicative of the hydrocarbon contamination adsorbed on the top of niobium oxide surface. The large hydrocarbon contamination in CS polished samples (P78C and P79C) could be related to the presence of organic components in the colloidal silica solution itself, whose composition is proprietary. On the other hand subsequent HPR treatment (P78H, P79H) was found to re-introduce hydrocarbon contamination which was unexpected. Although, HPR processing on BCP treated samples (P78H, P79H) was carried out in class 100 clean room, still it is an extra step that exposes the top oxide layer to the atmosphere for longer period (~1hr) than the BCP treated samples (P78B, P79B). This might have led to increased hydrocarbon contamination on HPR treated surface. The role of hydrocarbons assumes significance in light of a recent study that proposes link between suppressed superconductivity in Nb and the presence of hydrocarbons near stabilized dislocation bundles [12].

The change in hydrocarbon contamination was also co-related with the OH$^-$/O$^-$ ratio. This ratio measured using TOF-SIMS has been employed in the past, as an indicator of organic contamination [43]. Higher OH$^-$/O$^-$ ratio supposedly leads to increased hydrocarbon adsorption and this ratio was found to be independent of SIMS matrix effect [43-44]. The intensity of $^{16}$O$^-$ being saturated, we have plotted the average values of OH$^-$/$^{18}$O$^-$ ratio of each treatment against their respective normalized $C_xH_y^{+/-}$, NbH$_x^{+/-}$, and NbO$_x^+$ fragments; of which a few are shown in figure 3. The average normalized counts of the ion fragments measured after each treatment, have been scaled up or down to fit the plot. Figure 3, clearly reveals that OH$^-$/$^{18}$O$^-$ ratio was found to be maximum for colloidal silica polished samples (P78C and P79C) followed by HPR treated samples (P78H and P79H) and finally BCP treated samples (P78B and P79B). Normalized fragments related to hydrocarbons also followed the same pattern.

So, $OH^-/^{18}O^-$ ratio conclusively establishes the results presented against table 3. Moreover, increasing hydrocarbon contamination leads to simultaneous reduction of the niobium oxide fraction. Additionally, the $OH^-/^{18}O^-$ ratio varies linearly with hydrocarbon contamination on the surface. Hence, it can be safely concluded that BCP treatment minimizes hydrocarbon contamination on the top surface of Nb.

### 3.1.2. Comparison of Ionic fragments related to Niobium oxides:

Table 3 and figure 3 suggests that decrease of hydrocarbon contamination after BCP treatment led to dramatic increase of $NbO_x^{+/-}$ signals. This was expected as the hydrocarbons are present on the top of niobium oxide layer. Hence, reduction of hydrocarbons automatically increases oxide signals and vice-versa. This was further confirmed by the decrease of $NbO_x^{+/-}$ signals after HPR treatment. The change in $NbO_xH_y^{+/-}$ counts were found to follow the trend exhibited by $NbO_x^{+/-}$ counts. The increase of $NbO_xH_y^{+/-}$ counts after BCP treatment are indicative of increased Niobium hydroxides. Presence of niobium hydroxides has been confirmed using XPS technique [45] and is related to the water adsorbed on the niobium oxides. It can also form by wet reactions (BCP, CSP, EP) during oxide layer formation. Increase of $NbO_xH_y^{+/-}$ counts after BCP also leads to higher $OH^-$ counts. However, $OH^-$ counts does depend on other factors like; amount of water adsorbed on the surface, presence of oxygen and hydrocarbons with –OH bonding. On the other hand $NbO_xH_y^{+/-}$ and OH counts reduce after HPR treatment. This is again directly related to the reduction of $NbO_x^{+/-}$ counts which in turn depends on the hydrocarbon contamination as explained earlier. But, between the two sets of treatments the increase of $NbO_x^{+/-}$ counts was higher in P78 series. The reason was found to be low $NbO_x^{+/-}$ counts on the P78C samples compared to P79C, whose reason was not clear. It however indicates variability between similar treatments.

### 3.1.3. Comparison of contamination related to acidic residuals and its reaction products:

The other major changes after BCP treatment were related to the substantial increase in the intensities of acidic fragments. Table 3 shows that BCP treatment on CS samples led to increase of all acidic fragments by at least 6 times (~500%). $F^-$ intensity was found to increase drastically by more than 1300 times; niobium oxyfluorides fragments by 120 times; $C_xF_y^+$, $CH_xF^{+/-}$ fragments by 60 times; $Cl^-$ by 30 times and S by more than 5 times after BCP treatment. $NbOF^-$, $NbO_2F^-$, $NbO_3F^-$, $NbO_2F_2^-$ fragments has been used in the past as SIMS signature for Niobium oxyfluorides [46] and are formed due to the reaction of $F^-$ with niobium pentoxide. This was the reason for dramatic increase in fluorine counts. $C_xF_y^+$, $CH_xF^{+/-}$ fragments were supposedly sputtered from reaction product of hydrocarbons with HF acid residues. The conclusions regarding its origin would be explained later. The increase of $Cl^-$ and $S^-$ after BCP treatment was surprising. However it should be realized that chlorine and sulphur are always present as impurity in the BCP acid mixture in the form of HCl and $H_2SO_4$. Similar conclusions associated with increase of $Cl^-$ due to HF acid treatment on Si have been proposed earlier [47]. The reduction of $Cl^-$ counts after HPR treatment, confirms the argument. But it should be noted that $Cl^-$ may also get affected by atmospheric exposure [48]. That was one of the reasons why $Cl^-$ counts were found to increase in few samples that were HPR treated followed by BCP treatment (not in case of P78, P79). HPR treatment, on the other hand, reduces the acidic contamination of the top layer of BCP treated samples (P78B, P79B) by dissolution of the acidic residues and niobium oxyfluorides. The reduction is close to 80% of the BCP treated values for all the acidic residuals and reaction products. Reduction of fluorine by more than 40% has been reported earlier using XPS techniques [26]. The reduction of the acidic contamination after HPR process was similar in both the samples (P78H and P79H) which establishes the effectiveness and repeatability of the HPR process after any chemical treatment on Nb. However, the increase of S⁻ after HPR indicates contamination of ultra-pure water by leaching/ erosion of sulphur containing compounds from the pipes' walls used for transfer of HPR water.

The conclusions regarding origin of $C_xF_y^+$, $CH_xF^{+/-}$ fragments were based on observations on a different sample (P89) that was BCP treated, but not properly rinsed. This sample was kept exposed to atmosphere for few days prior to its analysis in static SIMS mode. The ion images acquired on this sample exhibited intense localized counts of $F^-$, $H^+$, $C^+$, $C_xF_y^{+/-}$, $CHF^{+/-}$ as shown in figure 5. This indicates that improper rinse leads to non-homogeneous distribution of HF residues that can react with hydrocarbons during drying. Presence of such areas with increased C have been reported recently using SEM-EDX [32] and XPS [41] techniques after BCP and EP treatment respectively. But, the results of SEM-EDX did not reveal the presence of fluorine, since the contribution of fluorine signal might be difficult to detect. However, XPS technique, does reveal the presence of increased F and C in some areas. Since the intensity of these fragments were always found to increase after BCP treatment (P78B and P79B) so its presence on chemically treated surfaces was undisputed. However, a thorough rinsing minimizes the chances of intense localization of this contamination. Reduction of these fragments after HPR treatment confirms its origin to acidic residuals. This type of contamination can also explain the presence of carbon contamination observed after BCP treatment [15]. These fragments also suggests that total counts of $H^+$, $C^+$ is dependent not only on hydrocarbon fragments but also on amount of $CH_xF_y^+$ type fragments. The role of acidic impurities and hydrocarbons assumes importance with its suspected role as paramagnetic impurities, which is further confirmed by the high curie constant of BCP treated Nb samples [11].

3.1.4. Contamination related to Metallic impurities:

The comparison of the metallic impurities were also done after each treatment. Table 3 mentions the changes of few of those metallic contaminants. $Na^+$, $K^+$, $Li^+$, $Mn^+$ (not shown here), $Fe^+$ and $Zn^+$ that were present on CS polished samples (P78C and P79C), were found to decrease consistently after BCP treatment. High concentration of $Na^+$, $K^+$ and $Li^+$ on the surfaces of CS polished sample was related to its presence in the colloidal silica solution used for polishing of Nb samples. The reason for reduction of $Mn^+$ and $Fe^+$ after BCP treatment was however not clear. On the other hand, $Cu^+$ and $Si^+$ was found to increase after BCP treatment. Although $Si^+$ is a major constituent of colloidal silica, the increase of Si contamination in BCP treated sample could be related to the Si impurity present in HF acid. Similar reasoning can be applied to $Cu^+$, which is an impurity in the acids used in BCP treatment. However, the actual cause needs to be explored further. The other metallic impurities like $Al^+$, $Mg^+$, $Cr^+$ (not shown here), $Ca^+$ and $Ni^+$ did not to follow any specific trend after BCP treatment, which could be a source of variability in BCP treated cavities. On the other hand, HPR treatment on BCPed samples (P78H, P79H) led to increase of most of the metallic impurities. Mg, Mn (not shown here), Ca, Fe, Ni, Cu and Zn increased dramatically; while $Na^+$, $Si^+$ and $Cr^+$, increased marginally after HPR processing. These metallic contaminations in HP rinsed BCP samples (P78H and P79H) was found to originate from the ultra-pure water used during HPR treatment. But, resistivity of the water measured at the source was 18.2 MΩ, which confirms it as ultra-pure water. This indicates that high pressure water (85kg/cm²) erodes the surface of stainless steel pipelines and brass connector present in the HPR set up. Hence, erosion was responsible for contamination of UPW by Cr, Ni, Fe, Cu and Zn. Similar results related to increase of Cu and Zn impurities on Nb surfaces, during HPR has been reported earlier [26]. Water analysis of HPR water revealed 5 to 10 times higher Cu (24 - 54 ppb) when compared to the UPW at the source. Ca, Mg and Si also increased after HPR treatment, which are known impurities of water.

However, the source of Ca, Mg and Si impurities was not clear. Since, SIMS is a highly sensitive technique, it was able to detect the changes of metallic impurities that may not have been possible with other techniques.

3.1.5 Reproducuibilty studies of surface treatments:

Repeatability of SRF cavity performance is a matter of concern. So, reproducibility studies were carried out for assessing the homogeneity of contaminations between spots as well as within a spot after each treatment on both sets of sample (P78 and P79). The non-homogenity within a spot was analyzed by observing the images of secondary ion distribution of respective impurities. On the other hand the inter-spot variation shown in Figure 4 is expressed in terms of percent standard deviation (SD).

$$\text{Spot to spot variation of } A^- = \left[ \frac{\text{Standard Deviation}\left(\frac{A^-}{Nb^-}\right)_{BCP}}{\text{Average of}\left(\frac{A^-}{Nb^-}\right)_{BCP}} \right] \times 100$$

Figure 4 shows the spot to spot variation of impurities after each treatment, represented by the respective ion fragments of hydrocarbons, oxides, acidic residuals and metals. It is clear that non-homogenity of metallic impurities within a sample was highest, irrespective of the type of treatment. Similarly acidic impurities also had higher variability of 25- 30% among BCP treated and subsequent HP rinsed samples. But, the samples had much better homogeneity of oxide fragments and hydrocarbons, except for P79C where the variability of oxide fragments are marginally higher. However, the inter sample inconsistency between the two sets of samples P78 and P79 is greater in CS polished and HPR treated samples.

Additionally the variation within a spot was found to be maximum in HP rinsed BCP treated samples (P78H, P79H) followed by CS polished samples (P78C, P79C). HP rinsed BCP treated samples (P78H, P79H) had extremely non-homogeneous distribution of metallic as well as non-metallic contamination within a spot (not shown here). Closer inspection of the ion images of metallic impurities on sample P78H and P79H along with its corresponding SEM image indicate preferable clustering of metallic impurities near the grain boundary steps, pits and any other steps formed inside grain. Similar results were also obtained during depth profiling of metallic impurities as shown in figure 9(a). This is an important observation as all non-barrel polished cavities have grain boundary steps at the equator region even after electropolishing treatment. Additionally these metallic impurities have magnetic moments that can locally affect the superconducting properties. Surprisingly, BCP treated samples showed extremely homogeneous distribution of metallic impurities within a spot. These results clearly specify the factors responsible for variability after the specific treatments. These results are well supported by research on FE craters that reveal the presence of variety of metallic and acidic impurities detected using AES and EDX techniques [20].

It is clear from the above discussion that BCP treatment reduces the hydrocarbon and metallic contamination along with increasing $NbO_x^+/Nb^+$ ratio in the top layer of Nb, but, leads to extensive contamination by acidic residuals and its reaction products. On the other hand HPR treatment reduces the acidic residuals but re-introduces hydrocarbon contamination along with extensive metallic impurities. Metallic contamination after HPR treatment can be controlled with improved HPR set up. The results of the static SIMS analysis was found to successfully relate the contamination on the surface to the type of surface treatment.

3.2. Oxide depth profile

The thickness of the top oxide layer [7,16-17] along with the type of suboxides [18] at the Nb-oxide interface affect the results of SRF cavity in more than one way. So, it was essential to study the oxide layer in detail. Slow sputtering using static SIMS approach [33,36,42] was adopted for such analysis. Using the above approach the intensity of various niobium oxide fragments of P78C (CS polished), P78B (CS+BCP treated), P78H (CS+BCP+HPR treated) and P115U (BCP+UHV annealed) samples across the oxide layer are plotted in figure 5. Niobium is known to be covered by a top niobium pentoxide layer followed by sub oxides. To identify various oxides (like $Nb_2O_5$, $NbO_2$, $NbO$ etc.) using SIMS it is essential to develop SIMS signatures of various oxide fragments plotted after exposing bare Nb surface to various partial pressure of oxygen and comparing them with niobium oxide of known stoichiometry. Considering these complications, we limit our discussion to the overall oxide thickness and presence of sub-oxides that seems possible. Earlier SIMS studies has established, that if, sub-oxides are present then the counts of $Nb^+$ must be lower than $NbO^+$ counts near Nb bulk-oxide interface [37,49]. Another research suggests that intensities of $Nb^+$, $NbO^+$, and $NbO_2^+$ peaks should not vary too much just before the oxide layer is completely removed [33]. As seen from figure 6, none of the treatments were found to exhibit such pattern. For a clear picture on the effect of treatments on the oxide thickness, it was necessary to divided into oxide layer in to three zones (1, 2, 3) as shown in figure 6.

   a) Zone 1- This zone exhibits sharp decline of the hydrocarbon contamination which is represented in the figures by $C_4H_5^+$ counts. The fall of $C_4H_5^+$ counts is almost synonymous with the sharp rise of the intensities of various oxide fragments to their maximum values. Hence, the maximum of point $^{18}ONb^+$ was decided as the end of zone 1. $^{18}ONb^+$ was chosen, because $NbO^+$ counts were saturated in the oxide layer analysis of P78H and P78B. The thickness of the hydrocarbon layer was found to be approximately 1/10th of the total oxide thickness. This is equivalent 0.3 - 0.4 nm thickness which was calculated using data from section 3.3. The values match well with earlier results using XPS techniques [10, 45].

   b) Zone 2- It is manifested by slow decrease of the intensities of various oxide fragments. This zone was considered to extend up to the point where the intensities of oxide fragments started to fall rapidly. The sharp fall also corresponds to the maximum intensities of $Nb_2^+$ and $Nb_3^+$ (not shown). This zone may be considered to indicate the stoichiometry of niobium pentoxide.

   c) Zone 3- The final zone is defined by the end of the oxide layer which results in sharp fall of intensities of oxide fragments along with stabilization of $Nb^+$, $Nb_2^+$ and $Nb_3^+$ ions. It also coincides with the rise of the $NbH_x^+$ fragments (not shown here).

These profiles are very similar to what has been observed earlier on buffered electropolished sample [33]. But one major difference exists. The intensity of $NbO^+$ is higher than $Nb^+$ in the first 250 – 300seconds. The reason could be attributed to the difference in the cracking pattern [37] of the hydrocarbon mixed niobium-oxide zone with $Bi_1^+$ gun as against $Au^+$ ion used earlier. So, except for the thickness of the second zone and slightly different top zone of P115U, no exceptional difference in the oxide layer was found due to various treatments. This also reflects the similarity of oxide stoichiometry in zone 2 and zone 3, irrespective of surface treatments. Since presence of sub oxides could not be verified we would evaluate the oxide thickness on the basis of the second zone alone. The ratio of the thickness of the second zone between 4 treatments was found to be 1: 1.18: 1.5: 1.9:: P78B: P78H: P115U: P78C (1525: 1800: 2300: 2900) that was based on the time required for sputtering the second zone. This result indicate that BCP treatment leads to minimum oxide thickness while CS treated samples have double thickness. This would be further verified in depth profile analysis section of 3.3. In other words, the second zone was found to be a good measure of the oxide thickness.

## 3.3 Depth profile till penetration depth (~50 nm)

The total spectrum of impurities to be analyzed within penetration depth of Nb can be categorized into interstitials, acidic reaction products and metallic contamination. Interstitial impurities like O, H, C and N and metallic impurities primarily increases the residual resistance thereby affecting the cavity performance. The precipitates of niobium hydride are considered to be most important factor responsible for the undesirable Q-slope and quenching at comparatively lower accelerating fields [7, 50-51]. Oxygen impurities not only affect $T_c$ [25] but is also responsible for weakening the surface superconductivity [11,52]. Oxides can also harbor local magnetic moments [18] which can be detrimental for SRF cavities. Carbon, on the other hand is responsible for decrease of the work function of Nb which can therefore turn into sources of electron emission [15]. The role of N is however not so clear [19], but recent use of nitridation technique to improve cavity $Q_o$, has generated interest [53]. Metallic impurities in Nb can also be a source flux trapping that can indirectly increase residual resistance. With the above background, it is important to realize the effect of each treatment on the total spectrum of impurities in the penetration depth. The depth profiles obtained using Cs-1keV and $O_2^+$-1keV sputter gun are plotted in figure 7 and 11 respectively. The depth profiles in figure 7 compare bulk hydrogen, oxygen, carbon and niobium oxyfluorides among various treatments mentioned in table 2. Similarly the metallic impurities are compared in figure 11. The depth profile of hydrogen and oxygen are compared using NbH$^-$ and $^{18}O^-$ ion fragments. On the other hand C$^-$ and NbOF$^-$ are also plotted to compare carbon and niobium oxyfluoride contamination. It should be noted that hydrogen contamination produces many fragments like H$^-$, H$_2^-$ and Nb$_x$H$_y^-$ (where x=1-5, y=1-4 with decreasing y as x was increased). Any of these fragments can be used for comparison of hydrogen contamination. In a similar manner NbO$^-$, NbO$_2^-$, NbO$_3^-$ can be used for oxygen; NbC$^-$ for carbon and any oxyfluoride fragments mentioned in section 3.1 instead of NbOF$^-$. The results of depth profile were found to be similar irrespective of the ion fragment used. The counts of nitrogen in depth profile mode, were too less to extract any meaningful conclusion. Although NbN$^{+/-}$ fragment has been used by researchers [7] for comparison, but this method has a major drawback of mass interference with NbCH$_2^{+/-}$ fragment which should not be overlooked. The result presented here are limited to comparison only and are not absolute concentration.

### 3.3.1 Hydrogen in Nb:

Evaluation of hydrogen contamination has always been critical for SCRF applications. Insitu growth of isolated and large niobium hydride precipitates was observed during cool-down of high purity Nb samples [50] which limit cavity performance. Figure 7(a-inset) clearly suggests that NbH$^-$ contamination in bulk Nb is maximum in CS polished samples. The contamination in BCP and subsequent HPR treated samples were marginally lower (see inset of 7a). Marginally higher values in CS polished samples may result from higher density of dislocation generated due to the stresses of mechanical polishing. Higher density of dislocation could trap larger quantity of hydrogen. In fact mechanically polished Nb samples had largest hydrides precipitate according to an earlier research [50]. On the other hand, BCP and CS samples that were simultaneously HV degassed exhibit reduction in bulk hydrogen. But, the surprising observation was, the reduction of NbH$^-$ in sample P117 (CSP+HV degassed) was only 2.8 times compared to 5.68 times in P115 sample (BCP+HV degassed). This difference cannot be generalized in simple terms of H getting trapped under dislocations. Although it is agreed that prior to HV degassing, the density of dislocations in mechanically polished Nb sample (P117C) would be higher than standalone BCP treated sample (P115B). But, the long degassing treatment that involves stress relief, annealing and grain growth would have been sufficient to bring down the dislocation density to similar levels in both the HV degassed samples (P115U, P117U). However, a recent study using Raman spectroscopy has proposed the presence of impurity stabilized dislocation bundles that can trap high levels of C, H, which are resistant to disintegration even during 600˚C HV degassing treatment [12]. In fact, P117U sample (CSP+HV degassed) has very high concentration of O, C, metallic impurities and pits compared to samples treated by other techniques. Moreover, strong correlation between the concentrations of oxygen and hydrogen near the surface was reported for several metals including niobium [53-54]. This could explain the difference in hydrogen contamination between P115U (BCP+HV degassed) and P117U (CSP+HV degassed) samples. However, the problem needs further investigation in future.

Another interesting feature involving hydrogen evaluation using SIMS technique relates to the presence of intense high resolution peaks of $NbH^-$, $NbH_2^-$ $NbH_3^-$ $NbH_4^-$, $Nb_2H^-$, $Nb_2H_2^-$, $Nb_2H_3^-$, $Nb_3H^-$, $Nb_3H_2^-$, $Nb_4H^-$, $Nb_5H^-$ in the spectrum obtained from the depth profile in BCP, CS and HPR treated samples. Earlier SIMS studies have also reported the existence of niobium hydride fragments upto $NbH_4^-$ [38] and $NbH_5^-$ [7]. This suggests that bulk Nb has unlimited areas to hold hydrogen. TEM Diffraction patterns have indicated the presence of β and ε niobium hydrides during cooldown of Nb samples to liquid nitrogen temperature [29-30]. However, recent ABF and HAADF studies have confirmed presence of β-niobium hydride near room temperature after EP and BCP treatment, where H interstitials were shown to occupy tetrahedral site [29]. This has been independently concluded using ERDA technique [8]. These β hydrides have face centered orthorhombic structure-FCO. This tetrahedral arrangement with fco lattice has been considered as 1Nb atom surrounded by 4H atom or vice versa [56]. In fact considering this approach we find that the sputtered fragments of the form $Nb_4H^-$, $NbH_4^-$ correspond well with the above arrangement of 1Nb atom surrounded by 4H atom or vice versa. Moreover, HV degassing treatment was found to drastically lower the intensities of the mass fragments containing higher stoichiometric H or Nb than the lower ones. In other words the reduction of $NbH^-$ (lower stoichiometry 1:1) was only 5.68 compared to 50 for $NbH_4^-$ (higher stoichiometry 1:4) when the intensities were compared between HV degassed sample (P115U) and BCP treated sample (P78B or P79B). The ratio of few other fragments are mentioned in table 4 along with their respective spectrums in figure 8a. The mass spectrum in figure 8a was plotted after cumulative counts acquired from 300seconds (~60nm) of depth profile of sample P78B and P115U. This suggests that HV annealing is successful in reducing the probability of Nb surrounded by 4H atoms and vice versa, which means that the tetrahedral arrangement of H in Nb is disturbed. This can also be interpreted as reduction of the probability of formation of β-NbH precipitates. In fact absence of hydride precipitates during cool down of HV degassed Nb samples has been observed by Barkov and his team [50]. It can hence be suggested that the SIMS fragments sputtered using these analysis conditions can serve as an indicator of the lattice arrangement as well as concentration of H within Nb matrix, which in turn relates to the probabilities of hydride precipitation. Further confirmation of this possibility would happen by observing insitu the variations of $NbH_x^-$ fragment ratio during cool down to lower temperatures. The work is in progress and needs re-designing of sample holder.

To eliminate doubts regarding background H of the analysis chamber playing any role in the depth profiles, an Nb sample with low H intensity or varying H intensity was required. Such $NbH^-$ profile was observed in Nb thin film deposited on Si wafer which was used for sputter rate estimation. Figure 8(a) clearly shows that the NbH⁻ intensity in Nb thin film varied by 4 orders from $10^4$ counts on surface to zero counts near the Nb-Si interface. Similar hydrogen profile of Nb deposited on wafers has been observed earlier [57].

### 3.3.2. Oxygen in Nb:

The depth profile of oxygen in figure 7(b) clearly distinguish the top oxide layer by sharp rise and fall of $^{18}O^-$ intensity. Once the oxide layer is removed a stable Nb⁻ signal was observed. The depth at which the Nb⁻ signal reaches a constant value was used to determine the interface between the surface oxide film and the bulk niobium. Based on this, the oxide thickness was estimated to be 2.8±0.1nm for BCP treated sample, 3.7±0.6nm for HPR sample and 4.3±0.4nm for CS samples. The thickness of UHV sample could not be estimated correctly using the above method, because Nb intensity stabilized after ~15nm depth. This could be related to the higher and varying oxygen content within UHV degassed samples. However considering partial stabilization of Nb⁻ counts, the thickness was estimated to be 3.8nm. The thickness observed are on the lower side of the values reported in the literature which ranges from 2 - 6nm [10,14,22,27,28,31,33]. But, lower values are possible since the samples were loaded within 30 minutes of the final preparation step (except for P115U and P117U). Oxide thickness of ~3nm after BCP treatment has been reported using TEM imaging [31]. In fact our BCP treated samples that were analyzed after exposure to atmosphere for 2 days had an oxide thickness of 3.5±0.1nm. Hence, within the experimental conditions, BCP treated Nb has minimum oxide thickness, while CS samples have maximum thickness. Marginal increase in oxide thickness after HPR has been observed, to which we would come back later. In contradiction to earlier results stating non-uniformity of oxide thickness on Nb samples [33, 58], our BCP treated samples were extremely uniform. Moreover, the ratio of the oxide thickness after 4 different treatment is 1: 1.33:1.37: 1.56:: P78B: P78H: P115U: P78C which co-relates closely to the results obtained in section 3.2. Marginal variation of the ratio could be related to the presence of impurity deposits on the surface of HPR, HV and CS polished samples.

On the other hand, oxygen contamination was found to be ~10 times higher in bulk Nb of annealed sample (P115U, P117U) and ~5 times higher in CS polished samples compared to BCP treated sample. The increase of oxygen by ~10 times is higher than ~6 times observed earlier [7]. However, the increased oxygen content in P115U and P117U indicates dissolution of oxygen into the bulk during annealing, which is known to be detrimental to superconducting properties. But, higher oxygen content in CS polished samples was related to the presence of Si impurities inside the bulk niobium. These Si impurities were found to be surrounded by oxygen as revealed by the ion images (not shown here). The oxygen depth profile also indicated increased oxygen intensity at the interface between the oxide and bulk Nb, for HP rinsed BCP samples (P78H, P79H). The increased oxygen intensity at the interface is shown encircled in figure 7b. Since HPR can only clean the top surface, so, it is improbable to change the interface oxygen content. However, a closer inspection revealed that the increase of oxygen at interface, after HPR treatment, was a SIMS artifact. The artifact can be explained on the basis of results established in section 3.1.4 and 3.1.5, which indicated non-homogeneous clustering of metallic contamination on HPR treated samples (P78H and P79H). Similar results were again verified during depth profiling. Fig 9 shows the ion image of such an area along with its corresponding SEM image, with Cu as a metallic deposit on the surface. Figure 9a-9b shows the cumulative ion image (the total depth) of Cu⁻ and O⁻, while figure 9d-9e shows the cross-section image (across the depth) of the same location. Figure 9f shows the ion image overlay of Cu⁻ (pink) and O⁻ (blue). It clearly shows copper impurities are present over the oxide surface. Same observations were made for many metallic impurities on HPR treated surfaces.

Using this background, the exploded profile of the area was sketched in figure 10(A). It depicts the pristine surface of high pressure rinsed BCP treated sample (P78H or P79H). Orange colored deposit resembles copper metallic deposit as an example. It is clear from figure 10(A), that, when the $Bi_1^+$ gun analyses oxygen from oxide layers at portions 'a' and 'b', the oxygen at portion 'c' remains embedded under Cu or any other metallic deposit. So the oxygen at portion 'c' would show up in the depth profile when the oxide layer (~3nm) is partially or completely removed at location 'a' and 'b' by the sputter gun as shown in figure 10(B). So even though ~3 nm is removed, oxygen remains on the surface due to the location 'c'. This remaining oxygen, shows up as marginally higher intensity near the interface. This creates an illusion in the depth profile of HPR samples as seen in figure 7(b). This assumption was further confirmed when impurities that are trapped inside oxide layer (like C⁻, F⁻, P⁻, $PO_x^-$, $NbO_xF_y^-$ etc.) also exhibited similar depth profiles (see figure 7c, 7d for C⁻ and NbOF⁻ respecively) for HPR treated samples. Moreover, such SIMS artifact was found missing in areas analyzed without grain boundary steps, pits and substeps. Infact, the metallic deposits (Cu, in this case) itself show their presence inside Nb sample, which in reality is present on the surface only. In other words, greater is the thickness of the metallic deposit on the surface of Nb, deeper would be its presence shown during the depth profile of the metallic impurity. Based on this argument; it can be concluded that the height of these metallic deposits would be closely equivalent to the extended depth to which it shows in the depth profile in comparison to the BCP treated samples. The calculation of the heights would be done later as shown in figure 11. However, presence of metallic impurities inside flat grains could neither be denied nor confirmed. But, earlier research have indicated that the oxide thickness increases with increasing water dose during HPR [59]. Surprisingly, they have earlier reported the presence of higher Cu and Zn on their HPR treated Nb surface [26], which is similar to our results. So, the increased oxide thickness after HPR treatment is a matter of conjecture and requires further studies. Moreover, this type of artefact may go undetected in most of the surface analytical techniques. In fact the observations by early researcher indicating non-uniformity of oxide thickness [58] can also be explained on this basis of impurities on the top surface of Nb.

Therefore, it can concluded that the oxide thickness is uniform across most samples, but the presence of impurities leads to misleading conclusions on oxide serrations inside Nb. This also contradicts the theory of H diffusion via serrated oxygen [7]. However, the effect after LTB could be different and is a study under progress. The results also show TOFSIMS technique as an effective way to measure the oxide thickness versus time which can used to control thickness dependent superconducting properties.

### 3.3.3. Carbon in Nb:

Figure 7(c) clearly suggests that carbon impurities are highest on the surface owing to hydrocarbon contamination and decreases sharply in BCP treated and BCP+HPR treated samples. But, the bulk carbon were consistently higher in CS polished as well as HV degassed samples. High carbon in CS treated sample (P78C, P79C) was due to SiC particles embedded inside Nb, during rough polishing with SiC coated surfaces. This was further confirmed by high and stable intensity of Si (not shown here). On the other hand, the reason for high bulk carbon in P115U and P117U relates to the presence of carbonaceous ($CO_2$ gases) contamination in furnace during degassing treatment at high temperature. But, the interesting part relates to the bulk carbon in P117U (HV degassed CS polished sample) sample, which was 2 order higher than P115U (HV degassed BCP treated sample), even though both samples were annealed simultaneously. This ratio was almost similar to the carbon intensity in CS polished sample and BCP treated samples. Hence, higher intensity of carbon in HV annealed samples is not only dependent on the contamination from the furnace chamber but also on the initial carbon concentration in Nb samples. This also confirms that sequence of treatments has a role in deciding the final impurity distribution inside bulk Nb. The depth up to which the carbon diffused in P115U was close to 800- 900nm. There is a high probability that the extra carbon that is present in UHV degassed samples would tend to transform into niobium carbide, owing to its limited solid solubility in Nb [60]. The other possibility of existence of carbon could be near dislocations bundles as suggested earlier [12]. Another important conclusion from the above analysis was that there was no segregation of carbon at the grain boundaries due to any treatments except HPR treatment. The segregation after HPR treatment is more of an artefact due to metallic deposits.

### 3.3.4. Acidic contamination in Nb:

The role of acidic contamination in the oxides are not specifically mentioned for cavity results, but these contamination within the oxide layers and bulk Nb can affect the near surface superconductivity [11]. Figure 7d clearly indicates intense accumulation of niobium oxyfluorides inside the oxide layer of Nb after BCP treatment. Other oxyfluoride fragments that were found in BCP treated sample include $NbF^-$, $NbOF^-$, $NbOFH^-$, $NbO_2F^-$, $NbO_2FH^-$, $NbO_3F^-$ and $NbO_2F_2^-$ which are known to be SIMS fragments sputtering out from niobium oxyfluorides [46]. Niobium oxyfluorides are formed during reaction of the oxide layer with HF acid. The oxide layer was also found to be loaded with $PO_x^-$, $NO^-$, and $F^-$. Although presence of $PO_x^-$ after BCP has been reported [19], we could not find such mention regarding niobium oxyfluorides. The presence of these products inside the oxide layer suggests that, once the samples are taken out from the BCP solution, the oxide layer starts growing immediately. But due to the high viscosity of the acidic layer on the Nb surface, they adhere on the surface and allow the reactions to proceed. But, the fast growing oxide layer traps these reaction products. This is the reason that all acidic reaction products except $F^-$, are confined within the oxide layer of BCP treated samples. The reason for diffusion of fluorine beyond the penetration depth was not clear. The depth up to which fluorine was found in BCP treated samples was close to $800 - 900nm$ (not shown here) in contrast to 150nm that has been reported earlier [20]. It has been reported that BCP treated samples display magnetic moments that come deeper from the bulk Nb. This leads to high curie constant in BCP treated samples than EP treated samples [11]. These authors suspect the role of F among other impurities [11]. In fact our earlier work reports lesser F diffusion in EP treated samples [61]. F and P impurities were found to increase by 2 orders at the Nb surface after BCP treatment. HPR treatment was successful in reducing the acidic contamination on the top surface of Nb, but the contamination within oxide layer and bulk was similar to BCP treated samples. Marginally, high intensity of oxyfluoride, $PO_x^-$, $NO^-$ fragment at the interface of HPR treated sample was again due to misleading depth profile results that has been already explained in section 3.3.2. Finally, HV degassing treatment led to drastic reduction of the F content to the levels observed in CS polished samples.

### 3.3.5. Metallic impurities in Nb:

Metallic impurities in Nb increase residual resistance and leads to flux pinning. Hence, it is important to identify the treatments that increases metallic contamination inside the penetration depth. The metallic impurities found after applying the above treatments had a definite pattern depending on the surface treatment and its sequence.

Since the metallic impurity distribution was treatment specific, so two impurities, Fe and Cu, were selected for comparison. Four distinguishable features could be observed from the depth profiles shown in figure 11.

i.   The pattern of metallic contamination behaved very similar to the C depth profile. The ratio of increase of Fe and Cu between BCP treated samples and BCP+HV degassed (P115U) sample was almost similar to CS treated and CS polished+HV degassed (P117U) sample. Similar results were observed for Si, Na, Mg, Al, Ca and Ni. This futher confirms that the impurities present in Nb prior to HV degassing treatment has a role to play. Another source of impurities in bulk Nb of HV degassed samples could be related to the contaminations present in the furnace itself. Reports in literature have indicated rise of residual resistance in SCRF cavity after 600˚C HV annealing [7,23]. This was evident from the increase in the metallic, carbon and oxygen impurities after HV annealing treatment. Additionally, thermometry studies on HV degassed cavities had suggested that the hot spots were observed at areas where it existed prior to the HV degassing treatment.  This is also demonstrated by the above results.

ii.  Among all treatments, BCP treated samples had minimum metallic contamination including Fe and Cu. Moreover, these samples exhibited homogeneous distribution of metallic impurities unlike samples treated by other methods.

iii. The metallic impurities in CS polished samples had very high intensity near the top surface, followed by constantly decreasing values with depth. Apart from Cr and Zn, metallic contaminations like Li, Na, Mg, Al, K, Ca, Mn, Ni and Fe were almost 4 - 10 times higher within the first 5–10nm compared to BCP treated samples. Few of these impurities were found to be present till 40 nm depth. Si contamination in CS polished samples were found to be much higher and extended to as deep as 3 microns in few spots. Extensive metallic contamination relates to the various media used during mechanical polishing of the samples.

iv.  HPR treated BCP samples (P78H, P79H) show very high intensity of metallic impurities like Ca, Cr, Ni, Zn (not shown here), Fe and Cu at the surface and found to be present to an extended depth in comparison to BCP treated samples. Higher intensity of metallic impurities at the surface of HPR treated samples co-relate well with our results presented in section 3.1. However, presence of these metallic impurities to an extended depth does not present a true picture as explained earlier. Using the analogy presented in section 3.3.2, the heights of the metallic deposits were calculated. The difference between the actual depth of impurity and the artifact in HPR treated samples are represented by the arrow shown in figure 11. The heights of the deposits were 3-4 nm for Al, Cr and Mn; 10 - 15nm for Fe, Ni, Cu and Zn. These values co-relate well with the extended depth of $^{18}O^-$ and $NbOF^-$ of HPR samples shown figure 9b and 9d respectively. The values are approximate, since the sputter rates of these deposit would be different from sputter rates calculated for Nb. It is clear from the above results that erosion of metallic surfaces which comes in the path of high pressure water was primarily responsible for contamination of the HPR water, which in turn contaminated the Nb surfaces. All the above results on metallic impurities were based on the maximum detection limits of the TOFSIMS instrument.

In summary, it can be concluded that BCP treatment results in minimum metallic contamination along with homogeneous distribution. On the other hand the advantage of reduction of hydrogen contamination during UHV annealing was offset by the presence of increased metallic impurities. One of the recent cures applied for reducing the metallic contaminations from HV degassing treatment suggests closing the end flanges with end caps made of Nb foils [62]. Moreover, metallic impurities were non-homogeneously distributed in all samples except BCP treated samples. It should be emphasized that none of the HV degassed samples exhibited segregation of metallic/no-metallic impurities along the grain boundaries. However, our recent analysis of Nb samples degassed above 1000°C does exhibit segregation of few metallic impurities near grain boundaries. This result would be dealt in our future studies.

## 4. Conclusions

TOF-SIMS technique was extensively and successfully applied to compare the vast spectrum of impurities in high purity Nb samples after various treatment simulating SCRF cavity processing. A step wise impurity analysis procedure was implemented which include analysis of impurity distribution on top monolayer of Nb surface using Static SIMS technique, oxide layer investigation by slow sputtering technique and impurity distribution analysis within the penetration depth (~50nm) by dual beam dynamic TOF-SIMS technique. The results clearly indicate that hydrogen impurity levels inside the bulk Nb was highest in CS polished samples. BCP treated and subsequently HPR samples had marginally lower H contamination than CS polished samples. The reduction of hydrogen after HV degassing treatment not only depended on $H_2$ partial pressure, but also on the previous treatment. This treatment also affected larger reduction in bigger $NbH_x$ ion fragments like $NbH_4^-$, $Nb_4H^-$ than the smaller fragments like $NbH^-$. It was also observed that interstitial and metallic impurities found inside Nb after HV degassing treatment were related to the impurities left over by previous treatments. On the other hand oxygen contamination was highest in HV degassed samples, while oxide layer thickness was found to be minimum (~2.8nm) in BCP treated samples. The thickness of oxide layers of CS polished and HPR treated samples were thicker than BCP treated samples. The misleading conclusions on non-uniformity of oxide thickness was also explained on the basis of presence of foreign impurities. Presence of sub-oxides could not be confirmed after any of the treatments, using slow depth profile technique. Two independent methods were presented for evaluation of hydrocarbon contamination. BCP treated, samples were found to have minimum hydrocarbon and metallic contamination. But, BCP treatment led to extensive contamination of the oxide layer by acidic residuals and acidic reaction products. Fluorine contamination was found to be present beyond the penetration depth of BCP treated sample (800 - 900nm). HPR treatment on the other hand was successful in reducing the acidic contamination on the surface of niobium, but led to non-homogeneous distribution of metallic and non-metallic impurities of Nb surface. The non-homogenity of impurity distribution was also discussed. This study clearly demonstrates the sensitivity of various impurities to the type of treatment and its sequence. The next step of this study would use the current set of results to analyze the complex EP treatment followed by low temperature baking. Further study is also needed for quantification of various impurities.

Table 4: The normalized $NbH_x^-$ ratio between BCP treated sample and HV degassed sample. The ratios represent the no. of times the fragments are higher in BCP treated sample.

| $\frac{\left(\frac{H^-}{Nb^-}\right)_{BCP}}{\left(\frac{H^-}{Nb^-}\right)_{UHV}}$ | $\frac{\left(\frac{NbH^-}{Nb^-}\right)_{BCP}}{\left(\frac{NbH^-}{Nb^-}\right)_{UHV}}$ | $\frac{\left(\frac{NbH_2^-}{Nb^-}\right)_{BCP}}{\left(\frac{NbH_2^-}{Nb^-}\right)_{UHV}}$ | $\frac{\left(\frac{NbH_3^-}{Nb^-}\right)_{BCP}}{\left(\frac{NbH_3^-}{Nb^-}\right)_{UHV}}$ | $\frac{\left(\frac{NbH_4^-}{Nb^-}\right)_{BCP}}{\left(\frac{NbH_4^-}{Nb^-}\right)_{UHV}}$ |
|---|---|---|---|---|
| 3.64 | 5.68 | 10.45 | 17.0 | 44.12 |

Table 1: Chemical composition and grain size of the niobium used for our analysis

| Material Source | RRR | Grain size | Composition (wt. ppm) | | | | | | | |
|---|---|---|---|---|---|---|---|---|---|---|
| | | | H | C | N | O | Ta | Fe | Cr | W |
| M/s Plansee, GmbH | 495 | 80 μm | 8 | 10 | 10 | 40 | 83 | <6 | <0.1 | 1.07 |

Table 2: Treatment sequence of four Nb samples (P78, P79, P115, P117) used in present analysis

| Steps | Treatment and its used abbreviation | | Parameters/ Time | First Sample | Second Sample | Third sample | Fourth Sample |
|---|---|---|---|---|---|---|---|
| 1(a) | Colloidal Silica polish | CSP | 40 nm Silica solution | P78C | P79C | X | P117C |
| (b) | Ultrasonic cleaning in Micro-90 | -- | 30 minutes | | | | |
| (c) | Ultrasonic cleaning in ultra-pure water | UPW | 30 minutes | | | | |
| 2(a) | Buffer chemical polish | BCP | 1 hr at ~11˚C | P78B | P79B | P115B | X |
| (b) | Ultrasonic cleaning in ultra-pure water | UPW | 1 hr | | | | |
| 3. | High pressure rinsing | HPR | 10 min at 85 kg/cm² water pressure | P78H | P79H | X | X |
| 4. | UHV annealing | UHV | 600˚C for 10 hrs at <2x 10⁻⁶ mbar | X | X | P115U | P117U |

Table 3: Table presents the changes in normalized intensities of various positive and negative ion fragments after BCP and HPR treatments on CS polished samples. *Calculation based on Absolute counts.

| Normalized negative fragments | Percentage change from CS to BCP | | Percentage change from BCP to HPR | | Normalized positive fragments | Percentage change from CS to BCP | | Percentage change from BCP to HPR | |
|---|---|---|---|---|---|---|---|---|---|
| | P78C to P78B | P79C to P79B | P78B to P78H | P79B to P79H | | P78C to P78B | P79C to P79B | P78B to P78H | P79B to P79H |
| Hydrocarbon fragments in % | | | | | Hydrocarbon fragments in % | | | | |
| H⁻ | -67 | -62 | 18 | 35 | $C_2H_3^+$ | -49 | -37 | 33 | 38 |
| C⁻ | -48 | -42 | 12 | 27 | $C_2H_5^+$ | -62 | -48 | 33 | 37 |
| CH⁻ | -59 | -53 | 16 | 34 | $C_3H_5^+$ | -68 | -58 | 49 | 51 |
| $C_3H_2^-$ | -58 | -51 | 32 | 40 | $C_3H_7^+$ | -83 | -76 | 72 | 104 |
| $C_3H_3^-$ | -44 | -31 | 39 | 29 | $C_4H_7^+$ | -84 | -77 | 63 | 87 |
| $C_4H_3^-$ | -75 | -71 | 35 | 69 | NbH⁺ | -45 | -44 | 18 | 36 |
| NbH⁻ | -65 | -62 | 27 | 43 | CNb⁺ | -16 | -22 | 13 | 21 |
| CHNb⁻ | -50 | -49 | 30 | 41 | CHNb⁺ | -23 | -24 | 15 | 15 |
| Niobium oxide fragments in % | | | | | Niobium oxide fragments in % | | | | |
| O⁻ | 458 | 465 | -26 | -40 | NbO⁺ | 568 | 339 | -21 | -40 |
| OH⁻ | 93 | 106 | -3 | -14 | NbOH⁺ | 226 | 124 | -7 | -16 |
| $NbO_2^-$ | 1508 | 843 | -20 | -40 | $NbO_2^+$ | 1,431 | 884 | -37 | -59 |
| $NbO_2H^-$ | 268 | 183 | 6 | -13 | Acidic residual/ reaction product fragments in % | | | | |
| Acidic residual/ reaction product fragments in % | | | | | N⁺ | 324 | 255 | -34 | -32 |
| F⁻ | 194194 | 130087 | -87 | -87 | $CH_2F^+$ | 12,856 | 14,579 | -81 | -82 |
| CF⁻ | 6542 | 9429 | -80 | -82 | Metallic impurity fragments in % * | | | | |
| NO⁻ | 676 | 506 | -57 | -51 | Si⁺ | 4 | 140 | 54 | -17 |
| S⁻ | 116 | 52 | 239 | 372 | Ca⁺ | -50 | 166 | 8852 | 424 |
| Cl⁻ | 3311 | 4794 | -23 | -80 | Fe⁺ | -84 | -60 | 9083 | 860 |
| $PO_2^-$ | 8122 | 21052 | -86 | -89 | Ni⁺ | -22 | 1029 | 4923 | 94 |
| $NbO_2F^-$ | 12608 | 16336 | -91 | -93 | Cu⁺ | 28 | 79 | 58,634 | 37,451 |
| $NbO_2FH^-$ | 2633 | 4043 | -85 | -87 | Zn⁺ | -85 | -74 | 34,221 | 12,931 |

Table 4: The normalized $NbH_x^-$ ratio between BCP treated sample and HV degassed sample. The ratios represent the no. of times the fragments are higher in BCP treated sample.

| $\dfrac{\left(\frac{H}{Nb^-}\right)_{BCP}}{\left(\frac{H}{Nb^-}\right)_{UHV}}$ | $\dfrac{\left(\frac{NbH^-}{Nb^-}\right)_{BCP}}{\left(\frac{NbH^-}{Nb^-}\right)_{UHV}}$ | $\dfrac{\left(\frac{NbH_2^-}{Nb^-}\right)_{BCP}}{\left(\frac{NbH_2^-}{Nb^-}\right)_{UHV}}$ | $\dfrac{\left(\frac{NbH_3^-}{Nb^-}\right)_{BCP}}{\left(\frac{NbH_3^-}{Nb^-}\right)_{UHV}}$ | $\dfrac{\left(\frac{NbH_4^-}{Nb^-}\right)_{BCP}}{\left(\frac{NbH_4^-}{Nb^-}\right)_{UHV}}$ |
|---|---|---|---|---|
| 3.64 | 5.68 | 10.45 | 17.0 | 44.12 |

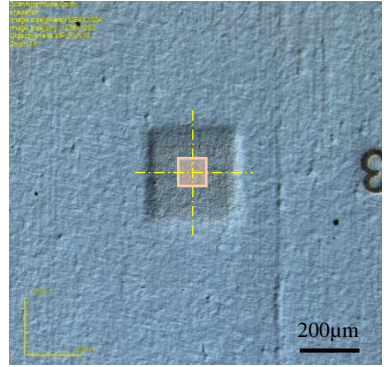

Figure 1: SIMS craters created by $Cs^+$ 1kV gun in CS polished Nb sample. Light brown box shows analysis area

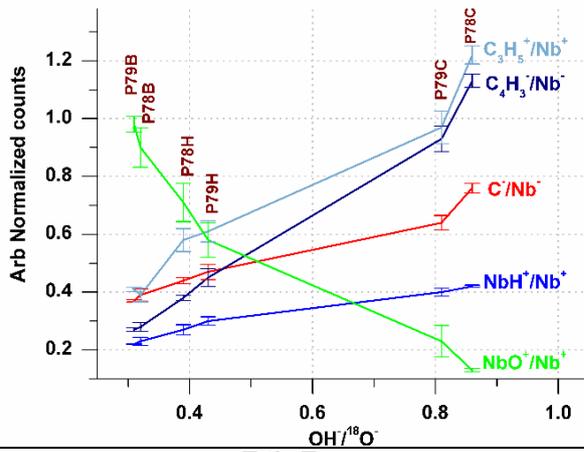

Figure 3: Variation of $OH^-/{}^{18}O^-$ ratio (x-axis) of each treatment plotted against variation of normalized hydrocarbon and oxide fragments (y-axis). The normalized values are scaled up or down to fit the plot.

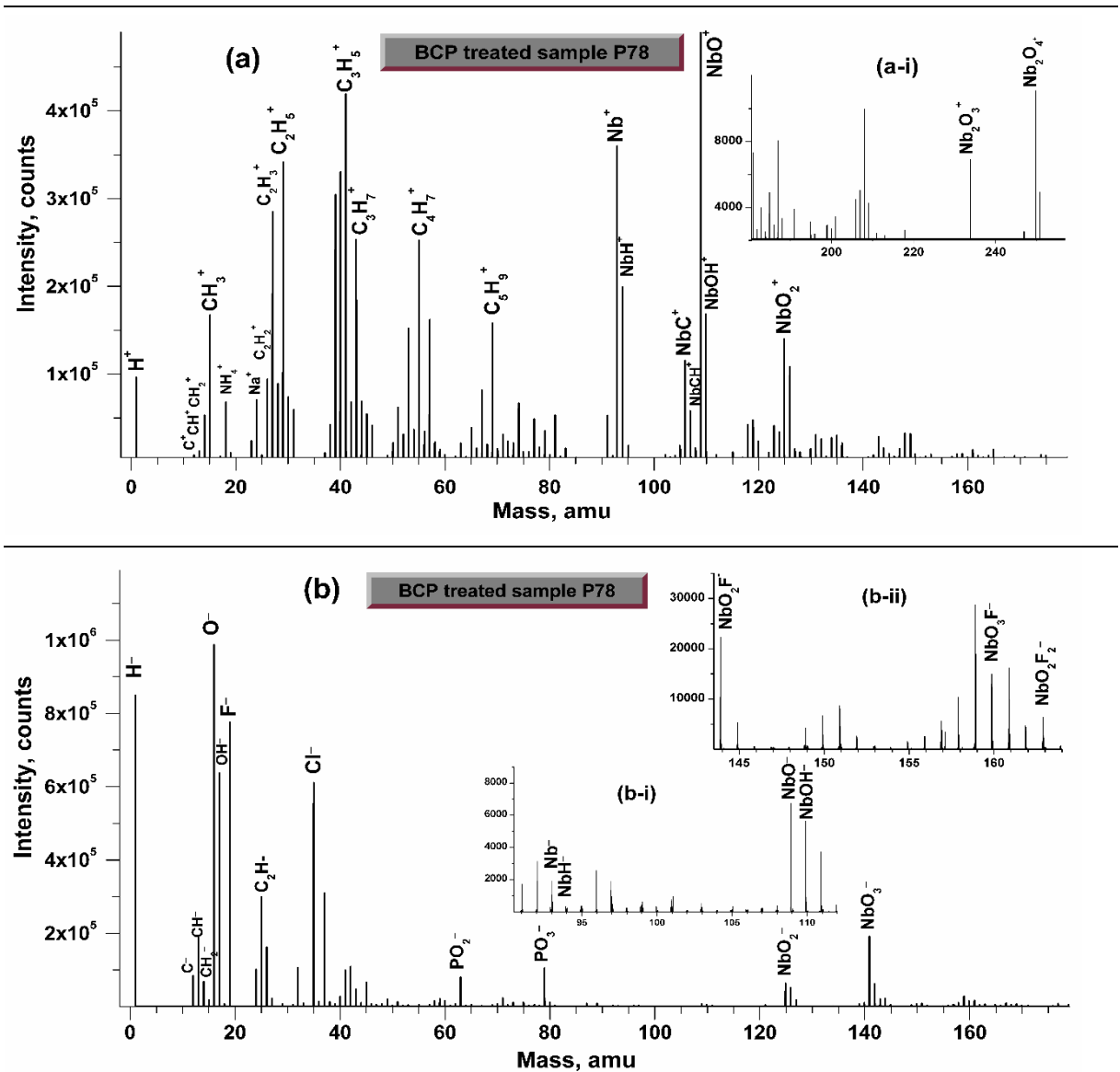

Figure 2: Typical positive (2a) and negative (2b) spectrum after 80 sec sputtering of top surface of P78B in static SIMS mode. Analysis gun Bi$_1$$^+$, 30keV, 2pA. Insets shows the exploded views of (2a-i) Nb oxide fragments, (2b-i) Nb, NbH$_x$ and NbO$_x$ frgaments and (2b-ii) Niobium oxyfluoride fragments.

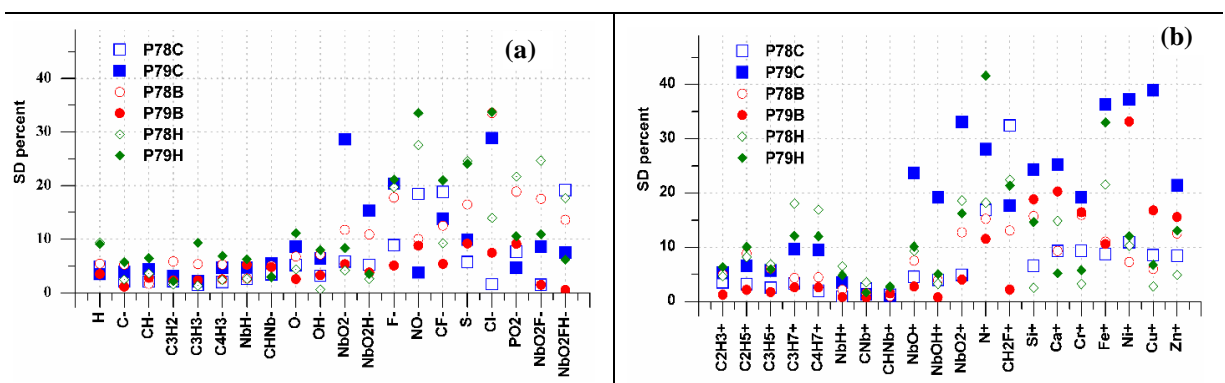

Figure 4: Spot to spot variation of (a) positive and (b) negative ion fragments after each treatment for P78C, P79C, P78B, P79B, P78H and P79H samples. The variation is represented in percentage. (See text for description)

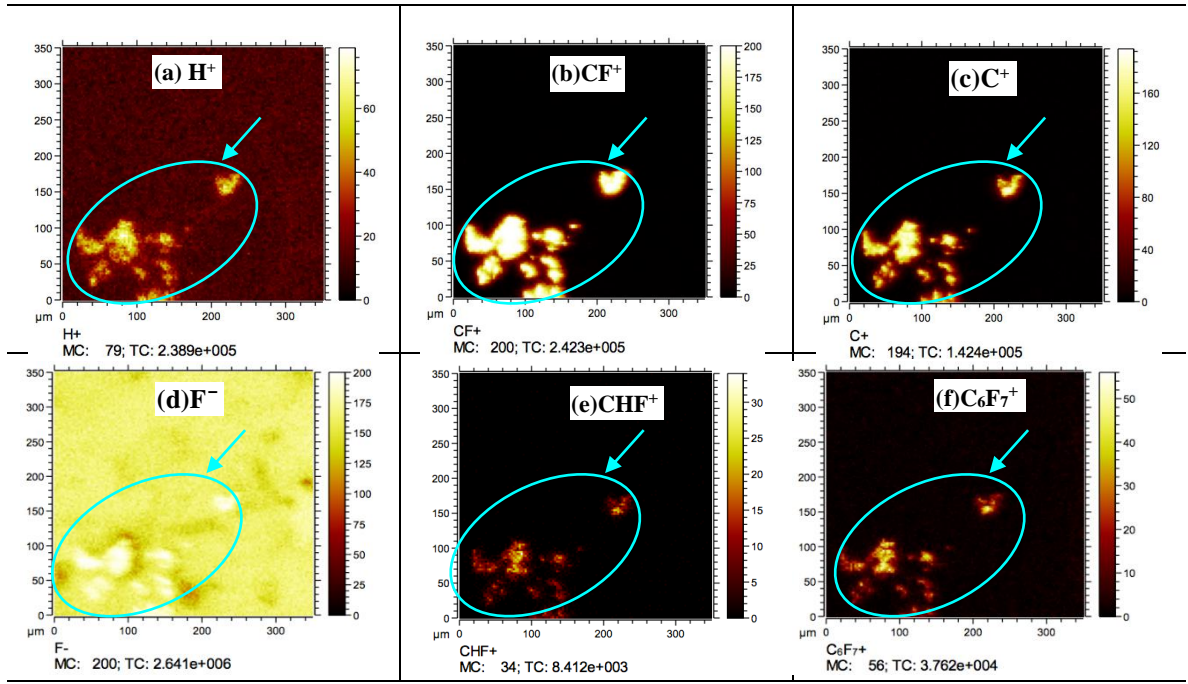

Figure 5: Ion images of BCP treated and marginally rinsed (P89) sample showing enhanced regions (circled) of (a) $H^+$, (b) $CF^+$, (c) $C^+$, (d) $F^-$, (e) $CHF^+$ and (f) $C_6F_7^+$. The bright yellow spots corresponds to maximum intensity of that particular ion; MC = Maximum count, TC= Total counts.

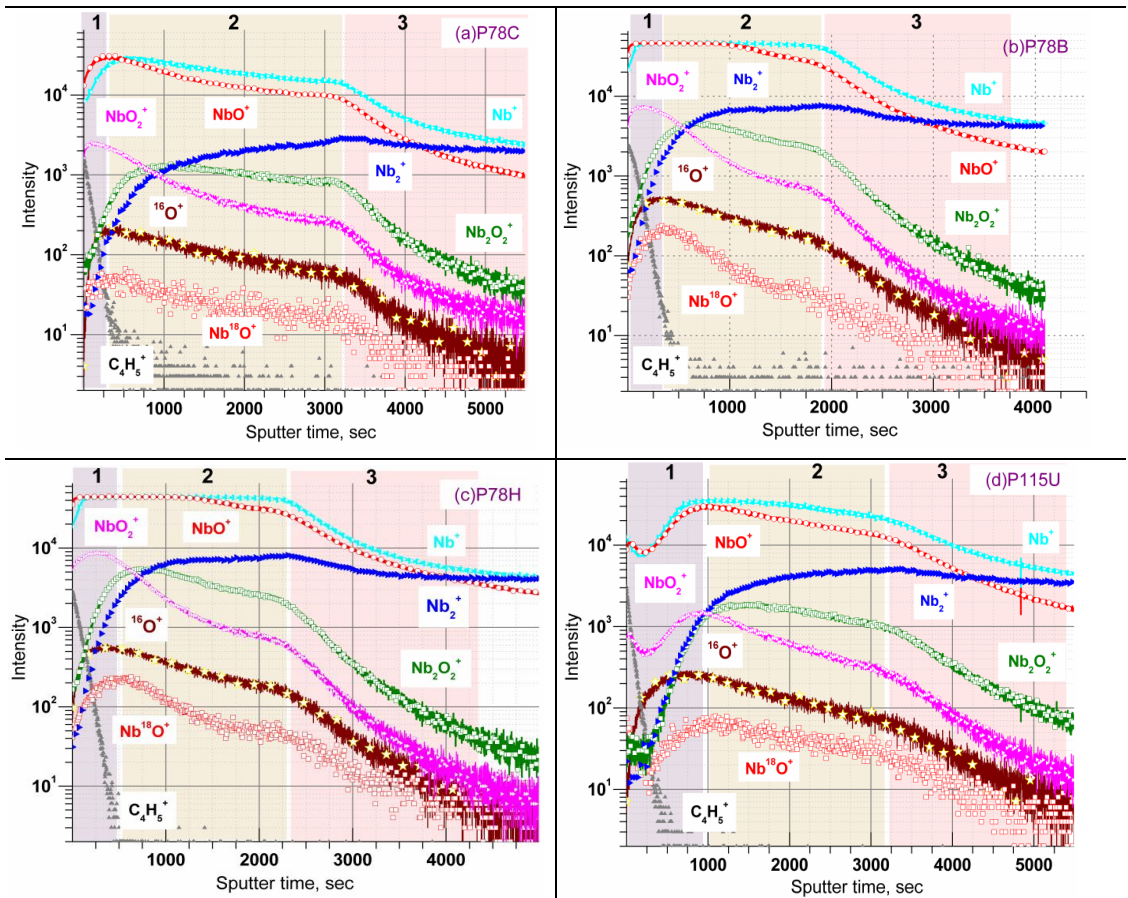

Figure 6: Slow depth profile of oxide layers of (a) CS polished- P78C, (b) CSP+BCP treated- P78B, (c) CSP+BCP+HPRinsed- P78H and (d) BCP+HV degassed- P115U samples.

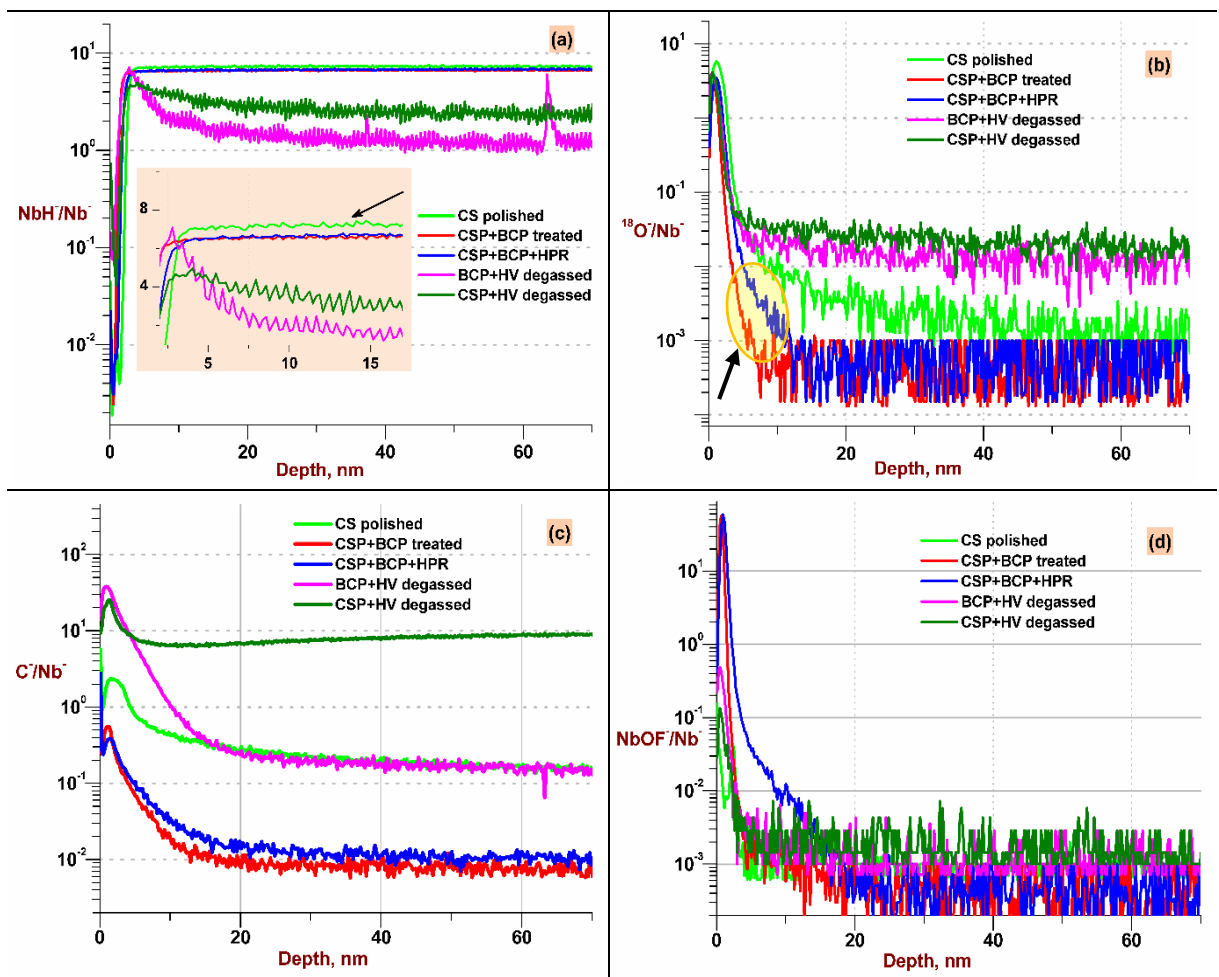

Figure 7: Depth profile of Normalized (a) NbH⁻, (b) ¹⁸O⁻, (c) C⁻ and (d) NbOF⁻ across the penetration depth of Nb prepared by CS polish, BCP, BCP+HPR, BCP+ HV, CS+HV annealing at 600°C for 10hrs. Y axis represents the normalized intensity, while X-axis represents the depth. Inset of figure 7a shows exploded profile of NbH⁻ near the Nb-oxide bulk interface. The depth profiles are average values of all the spots in all sets of sample.

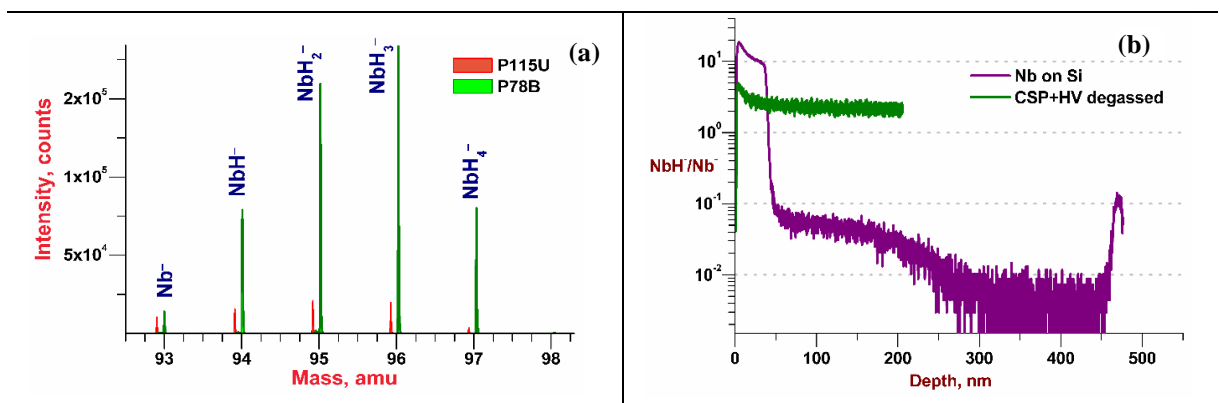

Figure 8: (a) The cumulative mass spectrum of NbHₓ⁻ (x=1-4) fragments acquired after 300sec of depth profile of P115U and P78B. The mass spectrum of P78B has been deliberately shifted to the right of P115U for comparison purpose alone (b) The variation of normalized NbH⁻ in Nb thin film deposited on Si wafer.

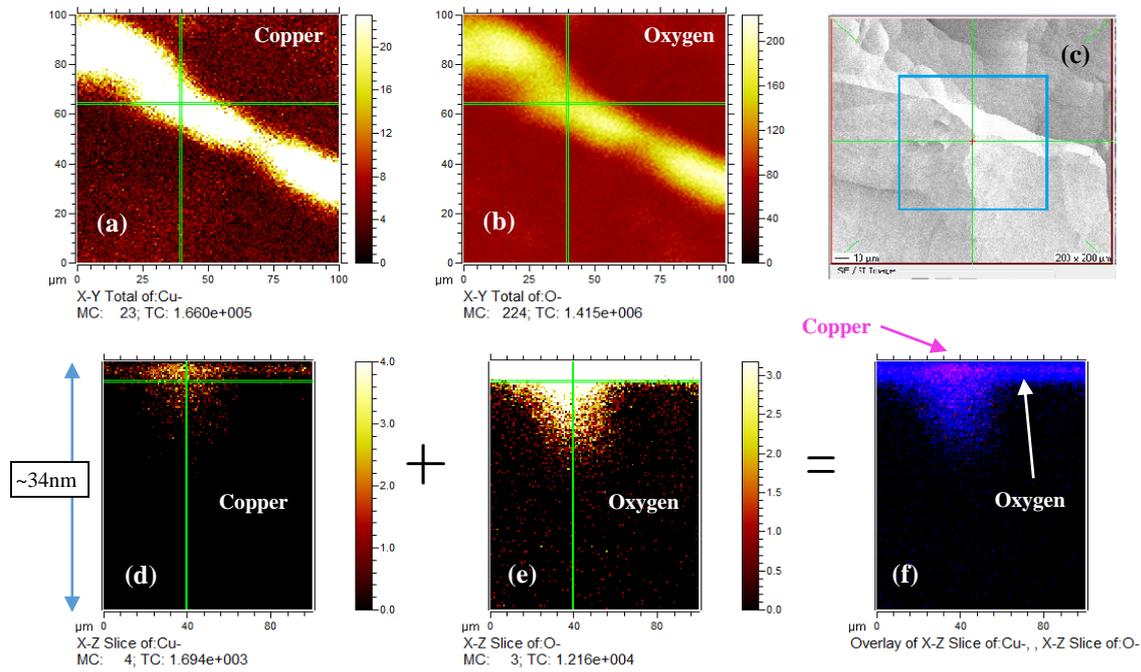

Figure 9: Secondary ion images of 100 µm² area of (a) Cu⁻, (b) O⁻, (c) The SEM image of the same spot , (d) cross section slice of Cu across the depth, (e) cross section slice of O⁻ across the depth (f) Pink and blue overlay of Cu⁻ and O⁻ from 9(d) and 9(e) across the grain boundary step in P79H sample after HPR. (9c). The counts of individual ion fragments in each pixel are shown in the intensity scale on the right of images. The bright yellow spots corresponds to maximum intensity of that particular ion; MC = Maximum count, TC= Total counts.

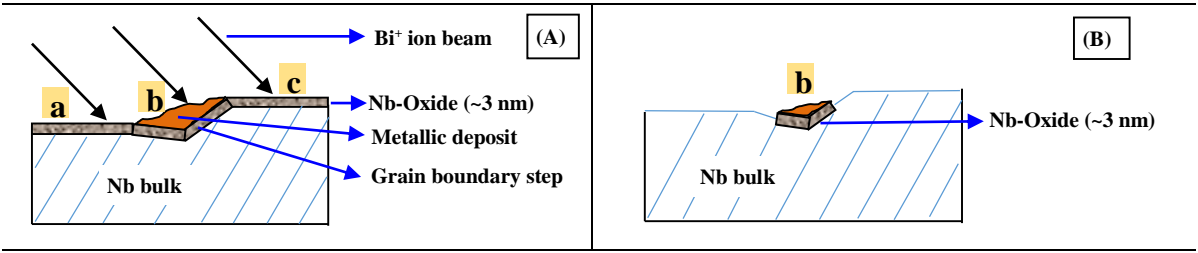

Figure 10: Sketch to represent the metallic deposits found on P78H, P79H as seen from ion images in figure 9. (A) Pristine Nb surface after HPR treatment; (B) After oxide layer is removed by sputter gun.

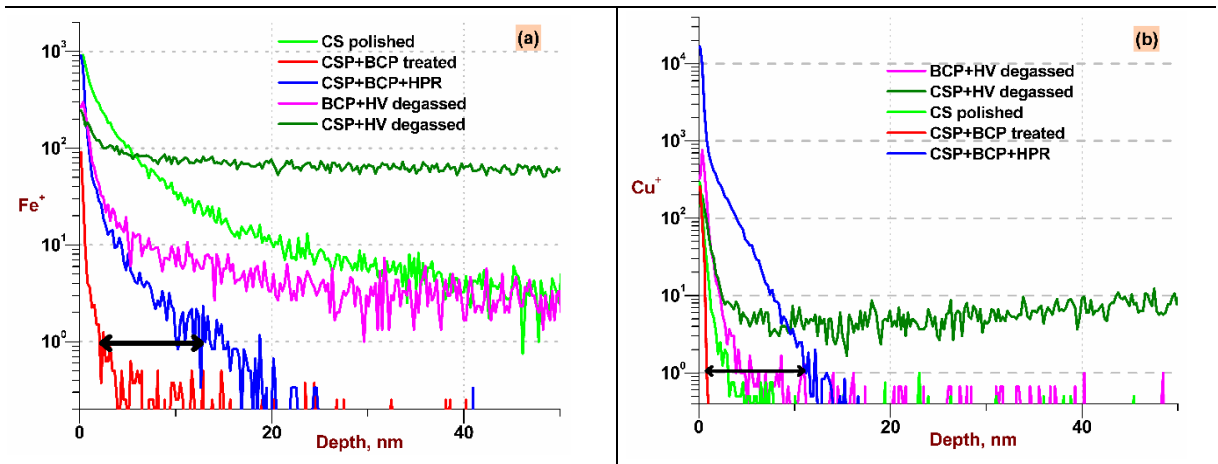

Figure 11: Depth profiles of (a) Fe$^+$ and (b) Cu$^+$ after the various treatments listed in table 2.